\documentstyle[prd,aps,preprint,epsf,epsfig,graphics,color]{revtex}
\firstfigfalse
\begin{document}
%\draft
\title{ 
%\begin{flushright} 
%{\small HLRZ-27/96}
%\end{flushright}
 The Phase Diagram of Compact QED Coupled to a Four-Fermi Interaction}
\vskip -1 truecm

\author{John~B.~Kogut }
\address {Physics Department, University of Illinois at Urbana-Champaign,
Urbana, IL 61801-30}

\author{Costas G. Strouthos}
\address{Department of Physics, University of Wales Swansea, Singleton Park, Swansea,
SA2 8PP, U. K.}

\date{\today}
\maketitle
\vskip -1 truecm

\begin{abstract}
Compact lattice Quantum Electrodynamics (QED) with four species of fermions is simulated 
with massless quarks by using the $\chi$QED scheme of adding a four-fermi interaction 
to the action. Simulations directly in the chiral limit of massless
quarks are done with high statistics on $8^4$, and $16^4$ lattices, and the phase
diagram, parameterized by the gauge and the four-fermi couplings, is mapped
out. The line of monopole condensation transitions is separate from the line of chiral
symmetry restoration. The simulation results indicate that the monopole condensation transition is
first order while the chiral transition is second order. The challenges in determining
the Universality class of the chiral transition are discussed. If the scaling region
for the chiral transition is sufficiently wide, the $16^4$ simulations predict critical indices far from mean
field values. We discuss a speculative scenario in which anti-screening provided by double-helix 
strands of monopole and anti-monopole loops are the agent that balances the screening of fermion 
anti-fermion pairs to produce an ultra-violet fixed point in the electric coupling.
\end{abstract}

\newpage

\section{Introduction}

Typical simulations of lattice field theories have grave difficulties determining features
of the theories' continuum limits. One needs accurate simulation data within the theory's scaling region
in order to measure critical indices with good control. Finite size effects must be monitored and 
either kept small so that infinite volume results can be read off finite volume data, or a series
of lattice volumes must be simulated so that finite size scaling can be used to determine scaling laws.

Carrying out such a program for fermion field theories is particularly daunting. Fermion 
algorithms typically require more computer time than purely bosonic models and dealing with almost massless
fermions, as must be done in the search for continuum limits of lattice models, is also very difficult.
However, it was noticed long ago that simulation studies of Nambu$-$Jona-Lasinio 
models tend to be much more quantitative
than those of other fermion field theories \cite{Looking}. In particular, the logarithmic
triviality of these models has been demonstrated, although determining logarithmic
singularities decorating mean field scaling laws is a
numerical challenge. The reason for this success lies in the fact that when one formulates
these four-fermi models in a fashion suitable for simulations, one introduces an auxiliary scalar
field $\sigma$ in order to write the fermion terms of the action as a quadratic form. In this
formulation $<\sigma>$, to be denoted $\sigma$ in much that follows,
then acts as a chiral order parameter which receives a vacuum
expectation value, proportional to the chiral condensate $<\bar\psi \psi>$, in
the chirally broken phase. Most importantly, the auxiliary scalar field $\sigma(x)$ acts as a site dependent
dynamical mass term in the fermion propagator. In the chiral symmetry broken phase,
the Dirac operator is now not singular for quarks with vanishing
bare mass and its inversion \cite{HMC}, \cite{HMD} is
accurate and very fast. The algorithm for Nambu$-$Jona-Lasinio models is
''smart'' --- it incorporates a potential feature of the solution of
the field theory, chiral symmetry breaking and a dynamical fermion mass, 
into the field configuration generator.

The good features of the simulation algorithm for the  Nambu$-$Jona-Lasinio model
can be generalized to lattice QCD \cite{KD} and noncompact QED \cite{PLB} by incorporating
a weak four-fermi term into their actions. These generalized models now depend on two couplings,
the familiar gauge coupling and a new four-fermi coupling. By choosing the four-fermi
coupling small we can be confident that all the dynamics resides in the gauge
and fermi fields and the four-fermi term just provides the framework for an improved algorithm
which allows us to simulate the chiral limit of massless quarks directly.

In the context of QCD, this approach is being used to determine the Universality Class of the
finite temperature transition of the isodoublet version of the model \cite{KD}. The method also has the
promise of producing light quark mass spectrum calculations of QCD, a physical situation which is not
practical in other approaches.

In the context of QED, this approach has been used to show the logarithmic triviality of textbook QED \cite{PLB}.
This is the noncompact version of the model, one free of monopoles and other topological
field configurations that a lattice model can support \cite{PLB}.

Our interest here lies in the compact version of the model, its magnetic monopoles
and how they might effect the model's continuum limit. The compact version of QED, Periodic QED, with dynamical fermions
has not been studied as well as the noncompact version of the model. We will be applying the standard algorithm,
the Hybrid Molecular Dynamics algorithm (HMD) \cite {HMD}, in the version in which the chiral limit of
massless quarks is accessible. This is the $\chi$QED formulation as reviewed above. This formulation is
particularly relevant because we are interested in a continuum theory with light physical fermions.

The Compact version of the four flavor QED has interactions coming from photons and magnetic monopoles coded
into the Periodic gauge fields. By adding a four-fermi interaction into this model which preserves a piece 
of the chiral symmetry of the massless model, we can study light fermions directly on the lattice and see how
fermion charge screening coming from light fermion loops effects the dynamics. The four-fermi interactions also
gives us the opportunity to separate the chiral transition of the model from its confinement/deconfinement
transition which is controlled by monopole condensation. These issues will be discussed more fully in the 
context of this paper.

It is crucial in all of this that fermion screening be accounted for accurately and realistically. It is
inappropriate to make any simplifications here because the character of fermion screening is not
understood outside perturbation theory. Some unbiased simulation studies are called for to see if, for example,
the anti-screening due to magnetic monopoles can balance the screening due to fermion loops and lead to 
ultra-violet stable fixed points \cite{vector}. This need further motivates us to consider the $\chi$QED
version of the model because then we are guaranteed that the fermions remain light even on the
strongly cutoff lattice, accounting quantitatively for fermion charge screening. In this formulation
of the model, there are light fermions even on the cutoff system, so we see the effects of fermion screening
throughout the entire calculation and not just in the subtle continuum limit.

Our lattice model has two couplings: the conventional gauge coupling of compact QED and the
four-fermi coupling of the Nambu Jona-Lasinio term. The uncoupled versions of Periodic QED and the
Nambu Jona-Lasinio models have been studied by lattice gauge theory methods and considerable quantitative
information is available to guide this study. We shall find two lines of transitions
in the two dimensional coupling constant parameter space of the compact U(1)-gauged
Nambu$-$Jona-Lasinio model. One line is associated with monopole condensation and the other with
chiral symmetry breaking. Long runs on fairly large lattices, $16^4$, indicate that the monopole
condensation transition is first order. A finite size scaling analysis of this transitions is necessary
to state this result with ''absolute'' certainty and such a study must await more computer resources. It is
well known that monopole condensation studies in compact lattice QED experience severe correlations in 
Monte Carlo time and very long runs are needed. In fact, many past simulations have been indecisive in
establishing the order of the transition and some studies have produced data that has
been fit with power behaved equations of state, characteristic of a second order transition. We will
show that our data cannot be fit sensibly with such a hypothesis and, in fact, a first order transition
where the monopole concentration jumps discontinuously to zero is strongly favored numerically.

The line of chiral transitions is distinct from the line of monopole transitions as long as the
four-fermi term is nonzero. One point along the line of chiral transitions
is studied in detail to determine the character of
the chiral transition. This work parallels recent studies of the noncompact version of the model
which concluded that the chiral transition was logarithmically trivial \cite{PLB}. Conventional ''wisdom'' would
suggest the same result here, since the monopole concentration vanishes on the line of chiral transitions.
However, this result is far from clear in our numerical work. The best fits to the chiral condensate and its associated
susceptibility are compatible with power laws of a second order transition, but the critical
indices are far from the expected mean field values. Unfortunately, the immediate vicinity of
the critical point cannot be well studied on this size lattice, $16^4$, because of finite size effects,
so the simulations reported here might not be registering the real continuum scaling laws of the
theory. Simulations on larger lattices closer to the critical point are required and studies
on $24^4$ and $32^4$ lattices are planned.

This paper is organized as follows. In the next section we present the formulation
of the lattice action and discuss its symmetries and general features. In the third section
we sketch the phase diagram and in the fourth section, the heart of the paper, we examine
several points in the phase diagram in detail and come to preliminary quantitative conclusions
about the first order character of the monopole condensation transition and the second order character
of the chiral transition. In the fifth section we comment on the puzzling nature of the 
results, present a physical picture of a nontrivial chiral transition in which anti-screening
due to entwined loops of magnetic monopoles and anti-monopoles balance screening due to fermion
anti-fermion pairs and suggest further research.

\section{Formulation}

To begin, consider the abelian-gauged Nambu Jona-Lasinio model with four
species of fermions. The Lagrangian for the continuum gauged Nambu$-$Jona-Lasinio model is,

\begin{equation}
L  = \bar \psi (i\gamma \partial -e \gamma A - m) \psi -
\frac{1}{2}G (\bar \psi \psi) ^2 -\frac {1}{4} F^2
\end{equation}

The Lagrangian has an electromagnetic interaction with continuous
chiral invariance
($\psi \rightarrow  e^{i\alpha \tau \gamma_5} \psi$, where $\tau$
is the appropriate flavor matrix) and
a four-fermi interaction with discrete ($Z_2$) 
chiral invariance
($\psi \rightarrow \gamma_5
\psi$). The mass term $m \bar\psi\psi$ breaks the chiral symmetries and will be set to zero
in much of the work that folows.
The pure Nambu$-$Jona-Lasinio model has been solved at large $N$ by
gap equation methods \cite{RWP}, and an accurate simulation study of it has been
presented \cite{Looking}. The discrete ($Z_2$) chiral invariant action produces a 
particularly efficient algorithm. Full chiral symmetry should be restored naturally
in the continuum limit in those regions of the paramter space where the four-fermi 
term proves to be irrelevant. The action with just the $Z_2$ chiral symmetry is
preferable over models with continuous chiral symmetry because they are not as easily and efficiently
simulated due to massless modes in the strongly cutoff theory.

It is useful to introduce an auxiliary random field $\sigma$ by adding
$-\frac{G}{2} ((\bar\psi\psi) -\frac{ \sigma}{G})^2$ to the
Lagrangian. This makes the Lagrangian a quadratic form in the fermion
field so it can be analyzed and simulated by conventional methods.
The model is then discretized by using staggered fermions.

The lattice Action reads, in the case where the gauge symmetry is interpreted
as a compact local $U(1)$ symmetry, following Wilson's original proposal \cite{wilson}:

\begin{equation}
S  =  \sum_{x,y} \bar\psi(x) (M_{xy} + D_{xy}) \psi(y) +
  \frac {1}{2 G} \sum_{\tilde x} \sigma ^2 (\tilde x) +
  \frac{1}{2 e^2} \sum_{x,\mu,\nu} (1-\cos(F_{\mu\nu}(x)))
\end{equation}

\noindent
where 
\begin{eqnarray}
F_{\mu\nu}(x)& = &\theta_{\mu}(x) + \theta_{\nu}(x+\hat{\mu}) +
\theta_{-\mu}(x+\hat{\mu}+\hat{\nu}) + \theta_{-\nu}(x+\hat{\nu}) \\
M_{xy}& = & (m + \frac{1}{16} \sum_{<x,\tilde x>} \sigma( \tilde x))
\delta_{xy} \\
D_{xy} & = & \frac{1}{2} \sum_\mu \eta_{\mu} (x) (
 e^{i\theta_{\mu}(x)} \delta_{x+\hat{\mu}, y}
- e^{-i\theta_{\mu}(y)} \delta_{x-\hat{\mu}, y} )
\end{eqnarray}

\noindent
where $\sigma$ is an auxiliary scalar field defined on the sites of the dual
lattice $\tilde x$ \cite{CER}, and the symbol $<x,\tilde x>$ denotes
the set of the 16 lattice sites surrounding the direct site $x$.
The factors $e^{\pm i\theta_\mu}$ are the gauge connections and
$\eta_\mu(x)$ are the staggered phases, the lattice analogs of the
Dirac matrices. $\psi$ is a staggered fermion 
field and  $m$ is the bare fermion mass, which will be set to 0.
Note that the lattice expression for $F_{\mu\nu}$ is the circulation of
the lattice field $\theta_{\mu}$ around a closed plaquette, the gauge field couples to the
fermion field through compact phase factors to guarantee local gauge
invariance and $\cos F_{\mu\nu}$ enters the action to make it periodic.

It will often prove convenient
to parametrize results with the inverse of the four-fermi coupling, $\lambda \equiv 1/G$,
and the inverse of the square of the gauge coupling, $\beta \equiv 1/e^2$.

The global discrete chiral symmetry of the Action reads:

\begin{eqnarray}
\psi(x) & \rightarrow & (-1)^{x1+x2+x3+x4} \psi(x) \\
\bar \psi(x) & \rightarrow & -\bar \psi (x) (-1)^{x1+x2+x3+x4} \\
\sigma & \rightarrow & - \sigma.
\end{eqnarray}
where $(-1)^{x1+x2+x3+x4}$ is the lattice representation of $\gamma_5$.

As we mentioned, interesting limiting cases of the above Action are the pure $Z_2$ Nambu$-$Jona-Lasinio
model ($e=0$), which has a phase transition at $G \simeq 2$ \cite{Looking} and
the pure lattice PQED ($G=0$) limit, whose first order chiral phase transition is coincident with its
first order monopole condensation transition
near $\beta_e \equiv 1/e^2 \approx 1.00$ for four flavors \cite{Dag}.

The lattice simulation code is very similar to others in this program. The systematic step size errors,
those varying as $dt^2$, the discretization of the molecular dynamics evolution equations in
HMD ''time'' $t$, have been studied in the past and are understood \cite{Looking,PLB}. Taking $dt \leq 0.01$
produced chiral condensates whose systematic errors were considerably smaller than their statistical
errors. Other algorithmic problems, such as finite size effects, tunnelling, long correlation
times were monitored carefully in the runs and will be discussed below when appropriate.

\section{Phase Diagram. Small Lattice Survey.}

\begin{figure}
\begin{center}
\scalebox{.5}{ \input{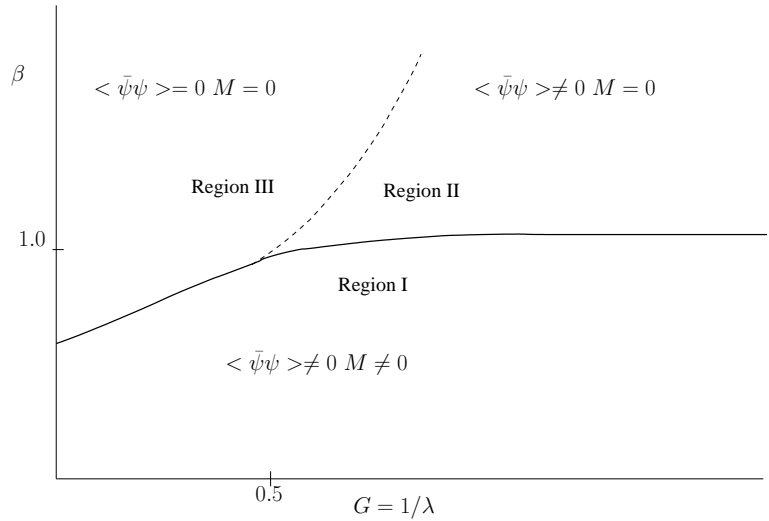} }
\caption{Phase Diagram of Gauged Compact $U(1)$ Nambu Jona-Lasinio Model.}
\label{fig:pqedz2}
\end{center}
\end{figure}

We scanned the 2 dimensional parameter space ($\beta$,$G$) using the Hybrid Molecular Dynamics
algorithm tuned for four continuum fermion species \cite{HMD}
and measured the chiral condensate, its susceptibility and the monopole concentration and its susceptibility as
a function of $\beta$ and $G$ on a $8^4$ lattice. These observables have been discussed extensively
in literature and we refer the reader to \cite{Dag,HW} for background.

For very small $G$ the chiral transition and the monopole condensation transition were coincident
to the accuracy and resolution of our survey. Abrupt jumps in the order parameters were measured and were interpreted as
signalling first order transitions. This is shown in  Fig. \ref{fig:pqedz2} as the thick line
of transitions extending to the $\beta$ axis. Earlier studies of the $G=0$ version of this lattice
model \cite{Dag} presented evidence for first order coincident chiral and monopole condensation
transitions on the axis in good agreement with this simulation.

As we follow the line of first order transitions into the phase diagram Fig.\ref{fig:pqedz2}, there comes a point
near $\beta \approx 1.0$ and $G \approx 0.5$ where the line of chiral transitions and the line of
monopole condensation transitions become clearly distinct. We will show measurements on the
vertical line $G = 0.70$ on a $16^4$ lattice which will make this point very clear. In particular, 
setting $G = 0.70$ and then increasing
$\beta$ in small steps from the strong coupling region of $\beta \ll 1.0$, we shall see good evidence
for an abrupt transition in the monopole concentration $M$ at $\beta \approx 0.95$. $M$ is large,
approximately $0.87$ at $\beta=0.953125$ and $G=0.70$, while it is small, approximately $0.018$ at
$\beta=0.9625$ and $G=0.70$. Throughout this region of the phase diagram, the chiral condensate $\sigma$ is
distinctly nonzero, so the system resides in the chirally broken phase even though $M$ vanishes. The
chiral order parameter $\sigma$ does, however, experience a jump from $\sigma \approx 0.55$ at
$\beta=0.953125$ and $G=0.70$ to $\sigma \approx 0.44$ at $\beta=0.9625$ and $G=0.70$, indicating that free
monopoles do contribute to chiral symmetry breaking, as expected, but the gauge coupling is sufficiently large
in this case that chiral symmetry breaking occurs even without the active participation of free monopoles. In fact
as we proceed up the vertical line $G = 0.70$, $\sigma$ falls only slowly and it is not until $\beta \approx 1.35(5)$
that it vanishes. The vanishing appears to be continuous, so the upper line of transitions,
the dashed line in the phase diagram Fig.\ref{fig:pqedz2},
appears to be second order. There is much more to discuss about the chiral transition and that will
occur below. However, it is eminently clear that the chiral transition occurs at a smaller gauge
coupling $e^2$ at fixed $G = 0.70$, and the two transitions, whatever their orders, are separate and distinct.

The dashed line of chiral symmetry breaking transitions continues to larger $G$ and higher $\beta$. The four-fermi
coupling alone breaks chiral symmetry at strong coupling, $G \approx 2.0$,
and the dashed line in Fig.\ref{fig:pqedz2} connects to that well-studied case
\cite{Looking}. Those studies of the pure four-fermi model were done with care and the logarithmic
triviality of the Nambu Jona-Lasinio model was confirmed by the numerical lattice simulation \cite{Looking}.

The first order line of monopole condensation transitions extends across the phase diagram
as shown in Fig. \ref{fig:pqedz2}. In fact, we will study it in detail by simulating the two-coupling
model on the horizontal line $\beta=1.0$ and variable $G$ and find evidence for a first order monopole
condensation line at $G \approx 1.8$ where $M$ jumps discontinuously to zero. We will see that $\sigma$ is
large and stable in the vicinity of this monopole condensation point, on both sides of it. Therefore, the
fermions have an effective mass comparable to the cutoff energy and should decouple from continuum physics,
if there is any, in the model. In any case, the model should not be significantly different from the pure
Compact Periodic lattice QED model and its monopole condensation should be essentially the same as in the pure gauge field
theory. The order of the monopole condensation transition has always been controversial in this case 
because of extraordinarily long correlation times. Our simulations show such effects, but our runs were 
long enough, we believe, that our configurations have achieved thermalization at each coupling and artificial
''rounding'' of the transition due to inadequate statistics has been avoided. We support the view that
the monopole condensation transition, which also signals the confinement-deconfinement transition in the pure
gauge theory case, is first order. Recent simulations of the pure Peridic QED model may be converging \cite{krauts}.

\section{$16^4$ Simulations of the Vertical $\lambda=1.4$ Line.}

Consider the observables, the chiral condensate and its susceptibility and the monopole concentration and its
susceptibility, as recorded in Table I. In this case the strength of the four-fermi interaction is set
to a constant $\lambda=1.4$ and the gauge coupling decreases as the simulation moves up the phase
diagram Fig. \ref{fig:pqedz2}. At strong gauge coupling, chiral symmetry is spontaneously broken and
and the monopole condensate is large and near its saturation value of unity. Both the chiral condensate and the monopole
concentrations are hardly fluctuating at $\beta$ in the vicinity of $0.90$ since their respective susceptibilities
are relatively tiny there. However, there is a clear jump in the monopole concentration to ''zero'' at
$\beta \approx 0.956$ and the chiral condensate experiences a discontinuity here also, but it does not vanish.
The chiral condensate will be discussed below. Here we concentrate on the monopole physics first.

\begin{figure}

\centerline{
\epsfxsize 4.0 in
\epsfysize 3.0 in
\epsfbox{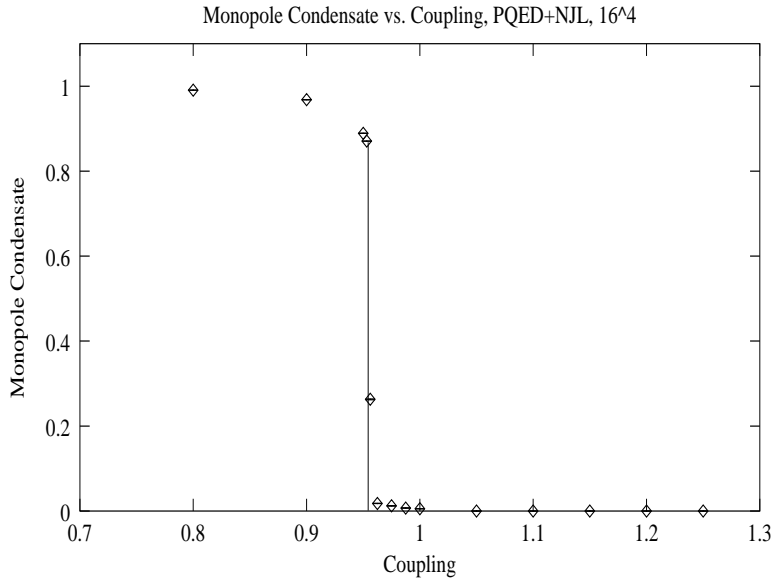}
}
\caption{Monopole concentration vs. gauge coupling $\beta$ at fixed four-fermi coupling $\lambda=1.4$.}
\label{fig:pqedM}
\end{figure}

Conventional wisdom would state that there is a confinement/deconfinement phase transition 
along the $\lambda = 1.4$ line at $\beta \approx 0.956$.
On the strong coupling side of this point there is a monopole condensate which causes confinement in the sense that
all the physical states are neutral and the electric charge is a good quantum number. In such a phase one expects
and finds chiral symmetry breaking. The character of the phase transition, whether it is first or second order, is crucial.
In Fig. \ref{fig:pqedM} we show the monopole concentration as a function of $\beta$ and suggest that there is
a discontinuity in the graph.

Attempts to fit the concentration data with a power law that would be indicative of a steep, but
continuous transition failed badly. For example, a power law fit of the standard variety,
$A (\beta_c-\beta)^{\beta_{mono}}$, to the points between $\beta=0.90$ and $\beta=0.95625$, had
a $\chi^2$ per degree of freedom in excess of $300$ and an exponent $\beta_{mono} < 0.025$. The
fit strongly suggests that a step function discontinuity is preferred as shown in the figure.

Why have other studies of this transition in the pure gauge theory produced power law fits? Since the effective
mass of the fermion is comparable to the cutoff in this part of the phase diagram, there should be no 
significant difference between the model simulated here and the simpler pure compact QED model. The
pure gauge part of the action is the same in both cases and the effect of the heavy fermions should be accommodated
by a tiny shift in the coupling constants. We noticed in our runs, however, a point discussed
extensively in the literature: extremely long relaxation times are needed to find the equilibrium
state of the pure compact model. If we had stopped our simulations with half the statistics we ended
with, there would have been some artificial ''rounding'' of the transition shown in the figure and a
power law fit may have been possible and misleading. Tens of thousands of HMD time units
are necessary to relax the system in the vicinity of the phase transition. Various authors have
suggested that since the model is expressible as the sum over closed loops of conserved monopole currents \cite{BMK}, local
Monte Carlo algorithms are particularly ill equipped to relax a system with such long range topological
structures constrained by global and local conservation laws.

Additional qualitative evidence for the first order character of the transition comes from the monopole susceptibility.
We see in the table that $\chi_M$ is highly spiked at the transition, jumping from $0.991(3)$ at $\beta=0.953125$ to
$27.6(3)$ at $\beta=0.95625$ and then falling to $16.52(4)$ on the weak coupling side of the transition at
$\beta=0.9625$. This suggests that the real susceptibility is a delta function in the large volume thermodynamic limit.
To really establish this claim, one would need a finite size study to monitor the size of the transition region and
the height of the susceptibility curves as a function of volume.

\begin{figure}

\centerline{
\epsfxsize 5.0 in
\epsfysize 4.0 in
\epsfbox{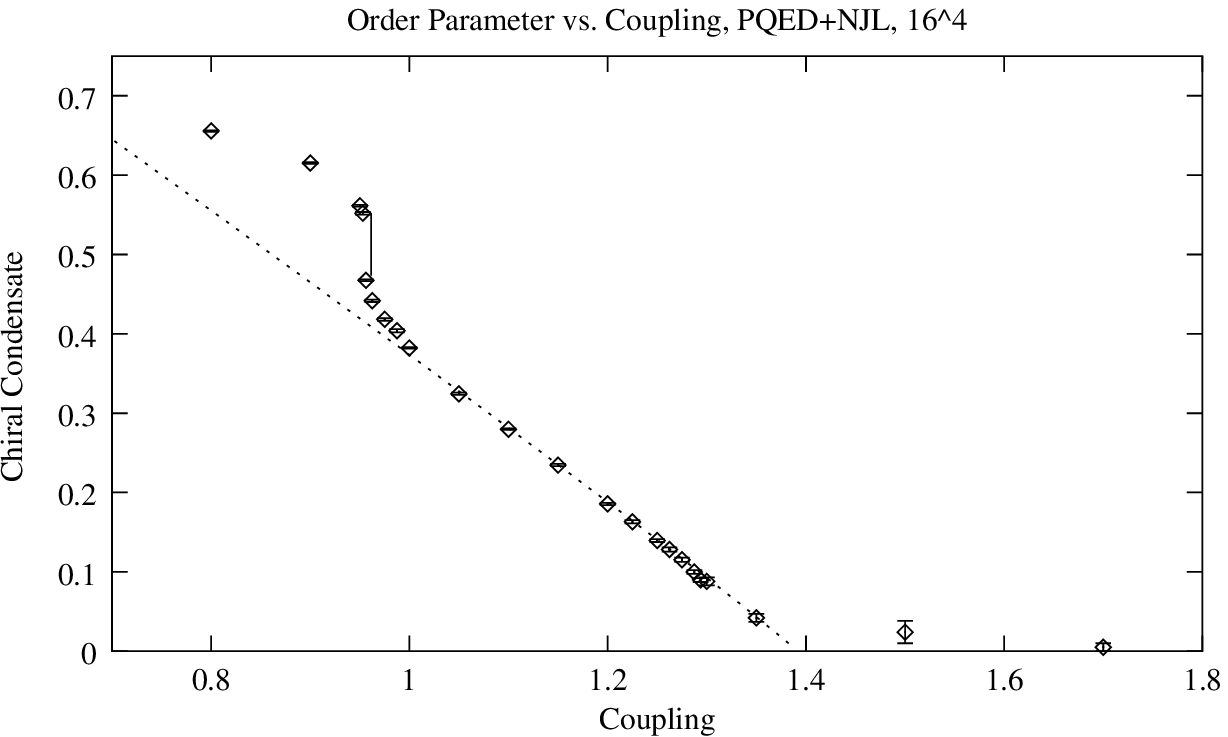}
}
\caption{Chiral condensate $\sigma$ vs. gauge coupling $\beta$ at fixed four-fermi coupling $\lambda=1.4$. The
dashed line is the power law fit discussed in the text.}
\label{fig:pqedpbp}
\end{figure}

Now consider the chiral transition along the $\lambda=1.4$ line. This is the major focus of this paper.
The raw data is shown in Table I. and is plotted in Fig. \ref{fig:pqedpbp}

At strong coupling chiral symmetry is strongly broken and the chiral condensate is large and
does not fluctuate severely. As in past studies, it proves convenient to monitor chiral symmetry
breaking through the vacuum expectation value of the auxiliary field $\sigma$ \cite {Looking,KD,PLB}.
The quark$-$anti-quark bilinear was also calculated in the simulation and is propotional to $\sigma$, in
accordance with the theory's equations of motion \cite{KD}.

The first interesting feature of this figure is the jump of the chiral condensate at $\beta=0.956$, at
the same point where the monopole concentration fell to zero. The fact that Fig. \ref{fig:pqedpbp} has
a jump at $\beta=0.956$ is further support for the first order character of the deconfinement
transition there. It is also interesting that the chiral condensate does not vanish on the weak coupling
side of the deconfinement transition. Apparently, the strong interactions in the theory are sufficient
to cause chiral symmetry breaking without confinement. We expect that the theory has a rich spectrum of bound states
on the weak coupling side of the transition at $\beta=0.956$ and $\lambda=1.4$, in accord with the
physical picture of chiral symmetry breaking in ref. \cite{casher}. It might be informative to
do spectrum calculations in this lattice theory in light of this result. It is very significant and
nontrivial that the chiral condensate does not vanish on the weak coupling side of the deconfinement transition
at $\beta=0.956$ and $\lambda=1.4$. Chiral symmetry breaking observed here is not due to the four-fermi interaction
alone. We know from ref. \cite{Looking} that a much stronger four-fermi interaction, $G \approx 2$, is needed to break
chiral symmetry in the absence of gauge interactions. In addition, simulations of the pure gauge theory
with four species of staggered fermions \cite{Dag} produced chiral condensates on the weak coupling side of the
transition which were consistent with zero. Physical mechanisms that could be causing substantial chiral
symmetry breaking in this region of the phase diagram will be discussed in the last section of this paper.

Returning to Fig.\ref{fig:pqedpbp}, we see that the chiral condensate falls essentially linearly in the
gauge coupling $\beta = 1/e^2$ until a chiral symmetry restoration transition is found in the vicinity of $\beta_c=1.393(1)$
and $\lambda=1.4$. As recorded in Table I., the Monte Carlo statistics in this region of the phase diagram
were quite considerable on this $16^4$ lattice. Nonetheless, as the critical point was approached, there 
were the usual troubles with with critical slowing down and tunnelling that hamper the predictive power of
these results. Critical slowing down causes the rising statistical error bars recorded in the Table
as $\beta$ approaches the transition. Tunnelling between $Z_2$ vacua also became
significant for $\beta \geq 1.29$ and limited our ability to simulate too close to the critical point. Notice that
the four-fermi term in the extended action is chosen to have a $Z_2$ chiral symmetry which is spontaneously
broken when $\sigma$ develops a vacuum expectation value. This symmetry was chosen for the four-fermi term
for several reasons. First, it is particularly easy to simulate and extract observables. In particular, as
long as the theory resides in a particular chiral vacuum, $\sigma$ and $\bar{\psi}\psi$ can be measured
directly, without the need for inserting a small bare fermion mass into the theory to pick out a unique
vacuum state. This is a great advantage because a bare fermion mass breaks chiral symmetry explicitly and leads
to ''rounding'' of the transition which makes quantitative studies of the transition and its Universality
Class very difficult. In addition, it is not necessary to measure observables like the absolute value of the
order parameter rather than the order parameter itself in order to simulate in a unique, specified vacuum state.
This is in contrast to using a $U(1) \times U(1)$ four-fermi term as is done in QCD applications of this
approach \cite{KD}. In that case the four-fermi term has a continuous chiral symmetry, so even in the broken phase the
order parameter $\sigma$ will average to zero as the phase of the vacuum state rotates over the unit circle. In the
QCD case, we use the magnitude of two auxiliary fields, $\sqrt{\sigma^2 +\pi^2}$, as an ''effective'' order parameter
to deal with this problem. Since fluctuations make the ''effective '' order parameter non-zero everywhere, rounding of the
transition is encountered. In the QCD case, it is important to use the $U(1) \times U(1)$ four-fermi term instead of
the $Z_2$ term used here, because one needs a massless pion in the system to have the correct light degrees 
of freedom characteristic of the  low energy spectroscopy of continuum QCD.
By contrast, the emphasis in QED is on the short distance behavior of the effective charge and other couplings,
and the low energy spectroscopy of the model is presumably irrelevant to the issue of the existence of the theory defined
at the ultra-violet stable fixed point.

The tunnelling between the two $Z_2$ vacua, $\sigma \rightarrow -\sigma$, limits our approach to the critical point
and appears to be the most damaging finite size effect we must deal with. Simulations are planned on larger 
lattices to lessen it. Of course, the simulations of this study on a $16^4$ lattice with high statistics  
represent a serious first step in this program.

Examining Table I., we note that the error bars in the $\sigma$ and $\chi$ data are particularly large for
$\beta$ between $1.30$ and $1.40$, reflecting the occasional, every thousand or so time intervals, tunnelling
between $Z_2$ vacua. The data for these points will not be used in the fits quoted here. Unfortunately, omitting
these data means that this simulation may not be sensitive to the real critical behavior of the model. For example, the
span of couplings from $\beta=1.1$ to $1.275$ used in the fits here might be outside the real critical region
of the theory, which might extend only between $1.30$ and $1.40$. It might be that we need data in this region to confirm
that this model has the ''expected'' logarithmically trivial scaling laws, like the non-compact model with four
fermi terms \cite{PLB}. This possibility and other issues of ''conventional wisdom'' will be reviewed in the section
on Conclusions below. At this point we will just do what we can do and plot and fit the order parameter and
susceptibility data where the finite size effects appear to be under control.

In Fig.\ref{fig:pqedpbp} we show the data on the chiral condensate fitted with a simple power law,
$\sigma = A(\beta_c-\beta)^{\beta_{mag}}$. The fit takes the data at $\beta$ ranging from $1.1$ through $1.275$ and
finds $\beta_{mag}=0.96(9)$ and $\beta_c=1.393(1)$. The confidence level of the fit is excellent, $89$ percent,
corresponding to a $\chi^2$ per degree of freedom of $1.12/4$. The central value for the magnetic critical exponent
$\beta_{mag}$ is unchanged by taking wider ranges of couplings and even approaching the apparent critical
point at $\beta_c=1.393(1)$ more closely, but the Confidence Levels deteriorate. We had expected a logarithmically
improved mean field fit here with the magnetic critical exponent near $1/2$, as was found in the non-compact
QED case in \cite{PLB}, but there is no sign of that behavior. A value of $\beta_{mag}=0.96(9)$ suggests a
nontrivial interacting theory, and is very perplexing, as will be discussed in the concluding section below.

\begin{figure}

\centerline{
\epsfxsize 4.0 in
\epsfysize 3.0 in
\epsfbox{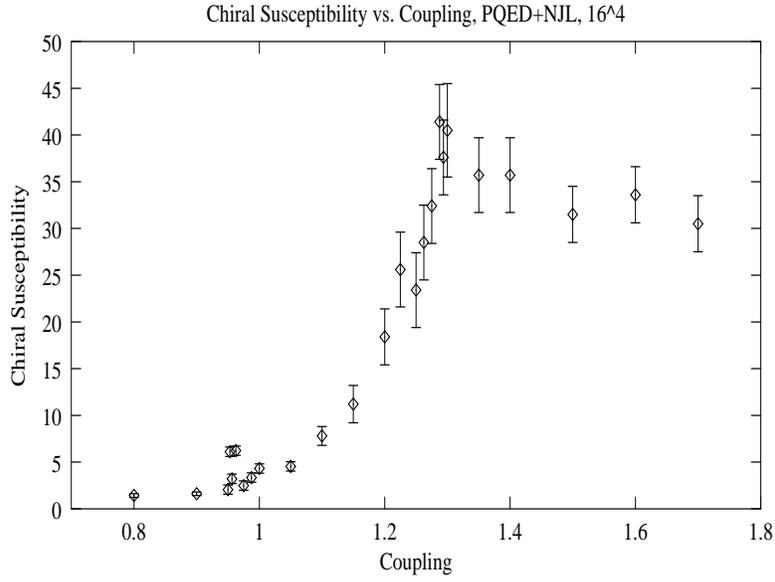}
}
\caption{Chiral susceptibility $\chi$ vs. gauge coupling $\beta$ at fixed four-fermi coupling $\lambda=1.4$.}
\label{fig:pqedchi}
\end{figure}

We also accummulated the fluctuations in the order parameter, the susceptibility $\chi$, as shown in Fig. \ref{fig:pqedchi}.
The values of $\chi$ in the immediate vicinity of $\beta_c=1.393(1)$ are certainly not reliable because of finite
size effects and tunnelling. However, the trend for the susceptibility to grow rapidly from $\beta=1.00$ to $1.275$
is clear and finite size effects appear to be under control on the $16^4$ lattice over this limited range of couplings.
Therefore, we attempted power law fits to the data and in the next figure, Fig. \ref{fig:pqedchiinv}, 
we show the reciprocal of the susceptibility, $\chi^{-1}$, fitted to the data over the range 
$\beta=1.05$ through $1.30$, $\chi^{-1}=B(\beta_c-\beta)^{\gamma}$.

\begin{figure}

\centerline{
\epsfxsize 5.0 in
\epsfysize 4.0 in
\epsfbox{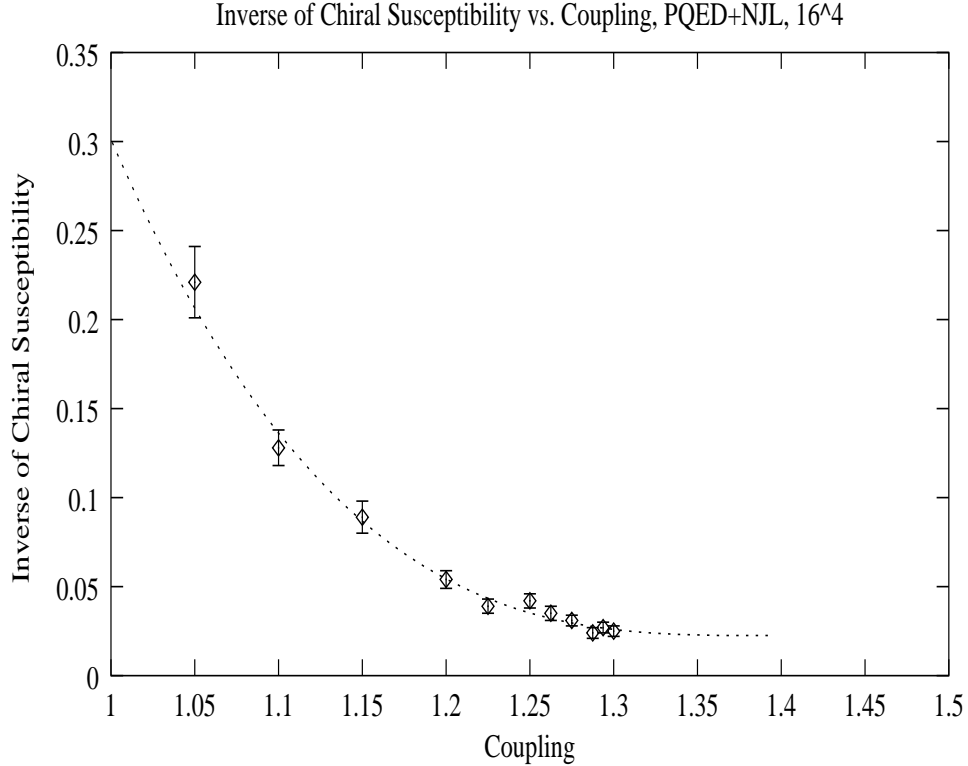}
}
\caption{Reciprocal of the chiral susceptibility $\chi$ vs. gauge coupling $\beta$ at fixed four-fermi coupling $\lambda=1.4$.
The dashed line is the power law fit discussed in the text.}
\label{fig:pqedchiinv}
\end{figure}

The fit is quite good, having Confidence Level of $48$ percent, $\chi^2$/d.o.f. $=7.5/8$. The critical index for 
the susceptibility is predicted to be $\gamma=3.1(3)$ and the critical coupling is again found to be $\beta_c = 1.393(1)$.
It is interesting that $\gamma$ is far from its mean field value of unity, although our reservations about this data are
the same as our reservations for the order parameter data and its fit.

The possible physical significance of this result will be discussed in the section on conclusions below. Since
susceptibilities typically are more sensitive to finite size effects than the order parameter, it would be
particularly informative to repeat this simulation on a larger lattice and attempt to track the
height of the susceptibility peak as a function of lattice size to determine $\gamma/\nu$, where $\nu$ is the
correlation length exponent, by finite size scaling methods.

\section{$16^4$ Simulations of the Horizontal $\beta=1.00$ Line.}

We also investigated the character of the monopole condensation transition along the horizontal line in the
phase diagram Fig. \ref{fig:pqedz2}.

\begin{figure}

\centerline{
\epsfxsize 4.0 in
\epsfysize 3.0 in
\epsfbox{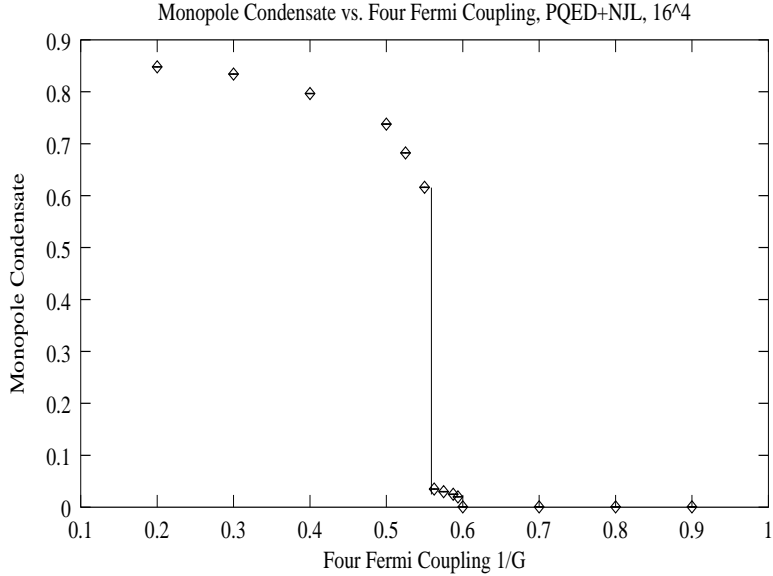}
}
\caption{Monopole concentration vs. gauge coupling $\beta$ at fixed gauge coupling $\beta=1.00$.}
\label{fig:mon16b1.0}
\end{figure}

We set $\beta=1.00$, on the basis of past studies of the pure gauge theory, and scanned in $\lambda$ until
we found the deconfinement transition. In fact, we started the simulations at strong four-fermi couplings,
large $G$ or small $\lambda= 1/G$, as given in Table II. For example, at $\lambda=0.20$, the monopole
concentration $M$ is large, $M = 0.8480(3)$ and its associated susceptibility is small $\chi_M = 1.24(1)$.
In addition, the chiral order parameter $\sigma$ is large, $2.909(1)$, at this parameter set and its susceptibility is
modest, $5.35(5)$, indicating that the fermions are irrelevant to the long distance dynamics in the model.
Throughout the entire $\lambda$ range, from $0.20$ through $0.90$ presented in the table on the horizontal line,
the fermions remain very massive. However, the monopole concentration experiences a deconfinement
transition near $\lambda = 0.5625$. As shown in Fig. \ref{fig:mon16b1.0}, the transition appears to be
a discontinuity, indicating a first order transition. In fact, power law fits to the monopole concentration
data near $\lambda=0.5625$ of the form, $C (\lambda_c-\lambda)^{\beta_{mono}}$ produced very small
indices, $\beta_{mono} < 0.10$ and very poor confidence levels, $\chi^2/$d.o.f. $>1000$. This result strongly suggests
that this transition is actually first order, with a step discontinuity, in agreement with the
results found at weaker four-fermi coupling, $\lambda=1.4$ and $\beta \approx 0.956$, discussed in the previous section.

Note from Table II. that huge statistics, more than $10,000$ Monte Carlo time units, were accummulated
near the transition to deal with the slow relaxation of monopole loops using a local, small change algorithm.

The fact that both horizontal and vertical scans of the confinement/deconfinement line give first order transitions
is a good test of our statistics and algorithm.

\section{Conclusions.}

When we began this study we believed that the monopole transitions would be very abrupt, perhaps first order,
and that the chiral transitions would be described by mean field theory, decorated by the logarithms of triviality,
as found in the noncompact theory, \cite{PLB}. Our expectations for the monopole transition held true, although it
took orders of magnitude more computer time to come to a decisive conclusion. We believed that
the chiral transition would be described by logarithmically improved mean field theory,
as found for the noncompact theory, because the monopole concentration would be vanishingly small in
the vicinity of the chiral transition, the periodicity of the action would become irrelevant, 
and the gauge field dynamics would reduce to photon exchanges,
just like in the non-compact model. Only if we could somehow tune the two couplings to that region of the
phase diagram where the first order confinement/deconfinement transition met the continuous chiral transition, did
we believe there was any possibility of interesting physics.

So we are left with a puzzling result: the chiral transition apppears to have critical indices far from mean field theory.
Just to illustrate and emphasize this point, assume hyperscaling and replace our numerical measurements of
the critical indices with integer predictions $\beta_{mag}=1$ and $\gamma=3$ they are consistent with. Then
the remaining critical indices would be $\nu=5/4$, $\eta=-2/5$, $\delta=11/4$ and $\alpha=-3$. Are such large deviations from
mean field behavior possible? When this model was studied in the limit of vanishing gauge couplings, toward the upper
end of the dashed line of chiral transitions shown in the phase diagram, Fig. \ref{fig:pqedz2}, the chiral transition was shown,
both analytically \cite{kocic}, and numerically \cite{Looking}, to be described by logarithmically improved
mean field theory. The algorithm used here but for vanishing gauge coupling gave results in fine agreement with
$1/N$ analyses. Even the exponents of the logarithms of triviality were compatible with theoretical expectations, even though
logarithms that decorate power law scaling laws are notoriously difficult to pin down. The extreme stability and accuracy of
the algorithm that uses the fermion dynamical mass to regulate it and guide it was cited as a reason for this numerical success.
If the ''results'' of this paper are correct, then we must conclude that the critical indices of the chiral transition
vary along the line of chiral transitions in Fig. \ref{fig:pqedz2} and the four-fermi interaction is not
irrelevant in this unfamiliar ground of strong coupling. Conventional wisdom, based on perturbation theory, 
would say that fermion vacuum polarization would
always screen the gauge couplings to zero leaving behind a Nambu Jona-Lasinio model that has no interactions in
a relativistic continuum limit.

What could the explanation of the simulation results of this study be? The obvious one is just that these results
are wrong, in the sense that they are not indicative of the true continuum limit of the theory. Perhaps if we could simulate
on larger lattices, closer to the critical point, we would find that the true critical behavior is, indeed, logarithmically
improved mean field theory as was found by exactly the same methods for the noncompact gauged Nambu Jona-Lasinio theory
\cite{PLB}. Perhaps the region in coupling in Fig.\ref{fig:pqedpbp} from $\lambda \approx 1.00$ to $1.3$
is outside the real scaling region and is strongly effected by irrelevant but large nonlinearities in
the Wilson Action for the compact $U(1)$ gauge fields. It may be that much larger lattices and much larger correlation 
lengths are needed to find the true continuum behavior in this model due to unusually large corrections to scaling for
this particular action.

We believe that these issues should be decided and our only tool that avoids uncontrollable approximations is numerical
simulations. This is a pity. Even numerical methods are sorely taxed by this problem. Nonetheless, simulations on
larger lattices, closer to the continuum limit are planned. Now that we know the interesting regions of the phase diagram,
we can focus in and, hopefully, get to the heart of the matter more efficiently than in this exploratory, but time consuming
study.

Early analytic studies of the gauged Nambu Jona-Lasinio model within a framework which included only 
ladder Feynman diagrams \cite{Love} and
which explicitly excluded fermion vacuum polarization predicted a line of nontrivial chiral transitions in 
the two-coupling phase diagram. It cannot be stressed too strongly, however, that this calculation was meant
as a model of Technicolor interactions and was not a solution of a field theory. It did not even include
those effects, fermion loops and vacuum polarization, that are expected to render the theory trivial.
However, other approximate approaches to this model which may account for screening to some degree have found a nontrivial
line of chiral transitions \cite{Az}. The reliability of this newer approximate approach is doubtful, however, because it
predicts, contrary to the simulation results of \cite{PLB}, that even the noncompact gauged Nambu Jona-Lasinio model
is nontrivial. The work reported here and in reference \cite{PLB} indicate that the noncompact model is
logarithmically trivial while the compact model may not be, contrary to \cite{Az}.

\begin{figure}

\centerline{
\epsfxsize 3.5 in
\epsfysize 2.5 in
\epsfbox{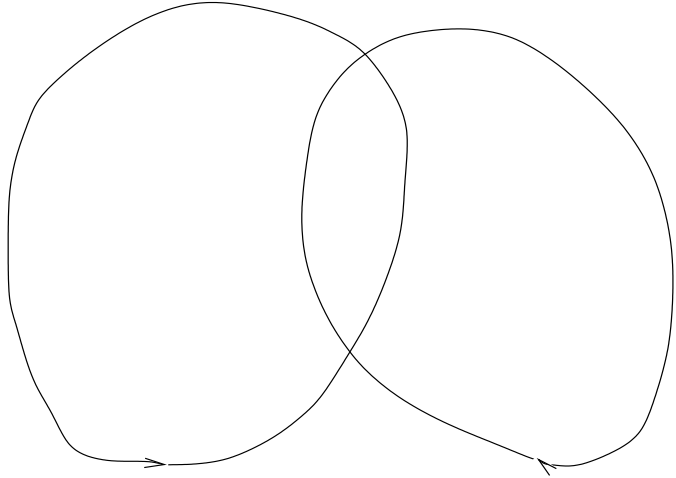}
}
\caption{Monopole loops in the confined phase, RNA.}
\label{fig:MMbarloops}
\end{figure}

\begin{figure}

\centerline{
\epsfxsize 3.5 in
\epsfysize 2.5 in
\epsfbox{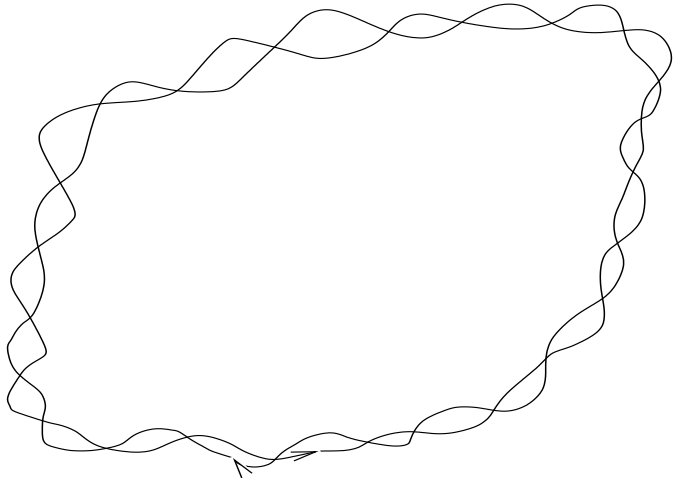}
}
\caption{Monopole anti-monopole helical loops in the deconfined phase, DNA.}
\label{fig:MMbarDNA}
\end{figure}

Let us end this discussion with some speculations which could guide the next generation of simulations
planned for this model. Suppose that the preliminary results presented here are basically correct. What sort of
physical excitations and interactions could support these results and how could they be discovered in the course
of a numerical study? The middle region of the phase diagram, region II in Fig. 1, needs clarification.
In this region $<\bar{\psi} \psi>$ is nonzero while the monopole concentration $M$ vanishes. The chiral transition
at the point $\beta_c=1.393(1)$ and $\lambda=1.4$ lies on the boundary between regions II and III in which
$<\bar{\psi} \psi>$ also vanishes.
The confinement/deconfinement transition at the point $\beta_c=0.956(1)$ and $\lambda=1.4$ lies on the boundary 
between regions II and I in which $M$ becomes nonzero. Conventional wisdom suggests that the confinement/deconfinement
transition between regions II and I is a four dimensional generalization of the Kosterlitz-Thouless
transition \cite{KT} which describes the two dimensional
planar spin model. In the two dimensional Kosterlitz-Thouless transition a state of vortex anti-vortex ''molecules''
ionize and form a plasma of vortices and anti-vortices. In four dimensional pure compact QED,
the transition from region II to region I should be driven by the ionization of strands of bound
monopole anti-monopole loops into a plasma of unbound individual loops of monopoles and loops of
anti-monopoles that cause confinement through the formation of electric flux tubes \cite{BMK}. 
For the lattice action used here, this transition
appears to be first order and does not lead to an interesting continuum field theory. A visualization
of random loops is shown in Fig.\ref{fig:MMbarloops} and it is tantalizing to call this the ''RNA'' phase of the theory,
and a visualization of double helix monopole anti-monopole strands is shown in Fig. \ref{fig:MMbarDNA}
and it is tantalizing to call this the ''DNA'' phase of the theory. Perhaps these pictures convey why it has proved
so difficult to decide the order of the confinement/deconfinement transition in compact QED. One must iterate
a local Monte Carlo alogorithm enough so that line singularities, not the point singularities of more
familiar transitions, can bind up into long, closed pairs. This is a very demanding requirement of such an
alogorithm, especially in four dimensions.

If the region II of the phase diagram really consists of double helix monopole anti-monopole strands, 
a ''DNA'' phase of compact QED, then we have a hint how the short distance properties of this theory
can be qualitatively different from noncompact QED. In particular, the monopole anti-monopole pairs,
which can exist in the compact model but not the noncompact one,
provide a medium which anti-screens electric charge. In classical electrodynamics, such an environment
raises the fundamental electric charge of an impurity $e^2$ to $\epsilon e^2$, where $\epsilon$ is the permittivity
of the vacuum. For a dilute background of monopole anti-monopole dipoles, each having a 
mean magnetic dipole moment of $\mu$, one finds that $\epsilon = 1+c\rho\mu^2$,
where $c$ is a positive constant and $\rho$ is the density of the magnetic dipoles. So, if these closed ''DNA''
strands remain relevant through region II until the critical point on
the boundary to region III is reached, there is a candidate mechanism in place for the
cancellation between the screening provided by the light fermion anti-fermion pairs and the anti-screening provided by the
''DNA'' strands of monopole anti-monopole pairs. As the critical point at $\beta_c=1.393(1)$ and $\lambda=1.4$ is
approached from region II, the fermions are becoming lighter ($\sigma$ is decreasing) and their contribution
to screening is increasing, thereby diminishing the renormalized electric charge. Finally, the renormalized electric
charge would be sufficiently small at the critical point that chiral symmetry would be restored 
and the theory would enter region III. So, the critical
point at $\beta_c=1.393(1)$ and $\lambda=1.4$ might be nontrivial because the
anti-screening due to the topology of the $U(1)$ gauge field action balances the screening due to
the fermion anti-fermion pairs of conventional vacuum polarization.

Other ideas concerning screening and anti-screening in compact gauge theories, such as
''collapse of the wavefunction'' and ''catalysis
of symmetry breaking'' \cite{DKK}, ideas inspired by monopole-induced proton decay \cite{Rubakov}, should be
considered in this framework again and might be ingredients in a successful quantitative implementation of the
monopole anti-monopole ''RNA-DNA'' scenario suggested here.

Apparently there is still much to learn in this difficult subject of strongly coupled gauge theories. Luckily,
advances in computer simulation power make many of these issues testable in the next round of investigations. In
particular, we plan measurements of the vacuum permittivity $\epsilon$, the renormalized electric charge and
the renormalized four-fermi coupling, the monopole anti-monopole spatial distribution in the vicinity of
an external charge etc. Measurements of the chiral condensate will be supplemented with measurements of the 
eigenvalue spectrum of the Dirac operator and the reliability of the $\chi$QED algorithm will be studied
in greater detail.

Perhaps some analytical progress can also be made. The Dirac quantization condition,
electric-magnetic duality transformations and other ingredients of Quantum Electro-Magneto Dynamics \cite{BS}
might be considered in this framework of chiral symmetry breaking.

\centerline{Acknowledgment}

JBK was partially supported by NSF under grant NSF-PHY01-02409 and CGS
was supported by the Leverhulme Trust grant.
The simulations were done at NPACI and NERSC. We especially thank
NERSC for their longterm, consistent support throughout this long ordeal.

%\clearpage

%\centerline{List of Tables}
%\vskip .5in

\narrowtext

\begin{table}
\begin{tabular} {cccccc}
$\beta$ & $\sigma$ &  $\chi_{\sigma}$   &  $M$   &  $\chi_M$   &  $Trajectories$   \\ \hline
$0.800$ &   0.6556(10)  &  1.43(20)      &  0.99067(2)   &  0.0451(2)   & 1500  \\
$0.900$ &   0.6153(10)  &  1.61(23)      &  0.96809(5)   &  0.1787(4)   & 2550  \\
$0.950$ &   0.5614(10)  &  2.05(40)      &  0.8890(2)    &  0.809(2)    & 2500  \\
$0.953125$ &0.5519(18)  &  6.11(50)      &  0.8708(2)    &  0.991(3)    & 3350  \\
$0.95625$ & 0.4674(10)  &  3.21(52)      &  0.263(1)     &  27.6(3)     & 4350  \\
$0.9625$ &  0.4416(15)  &  6.23(55)      &  0.0179(1)    &  16.52(4)    & 5100  \\
$0.975$ &   0.4182(18)  &  2.49(47)      &  0.0423(6)    &  16.0(1)     & 1000  \\
$0.9875$ &  0.4039(21)  &  3.35(50)      &  0.0661(5)    &  18.6(2)     & 1100  \\
$1.000$ &   0.3822(10)  &  4.32(53)      &  0.0214(1)    &  9.00(1)     & 8200  \\
$1.050$ &   0.3244(15)  &  4.52(37)      &  0.0001(1)    &  6.082(6)    & 2700  \\
$1.100$ &   0.2798(10)  &  7.80(56)      &  0.0001(1)    &  5.352(3)    & 14150 \\
$1.150$ &   0.2346(15)  & 11.2(1.0)       &  0.0001(1)    &  6.17(2)     & 6600  \\
$1.200$ &   0.1856(15)  & 18.37(1.6)      &  0.0001(1)    &  4.751(2)    & 10900 \\
$1.225$ &   0.1630(21)  & 25.58(2.4)      &  0.0001(1)    &  4.678(4)    & 7100  \\
$1.250$ &   0.1394(18)  & 23.41(2.1)      &  0.0001(1)    &  4.601(3)    & 8400  \\
$1.2625$ &  0.1281(27)  & 28.51(2.9)      &  0.0001(1)    &  4.568(4)    & 6000  \\
$1.275$ &   0.1153(27)  & 41.67(4.6)      &  0.0001(1)    &  4.539(4)    & 5200  \\
$1.2875$ &  0.0997(24)  & 37.03(3.7)      &  0.0001(1)    &  4.511(4)    & 4600  \\
$1.29375$ & 0.0897(25)  & 40.0(4.3)      &  0.0001(1)    &  4.501(5)    & 5400  \\
$1.300$ &   0.092(5)    & 50.0(6.5)      &  0.0001(1)    &  4.487(4)    & 5300  \\
$1.350$ &   0.042(5)    & 35.7(4.1)      &  0.0001(1)    &  4.413(6)    & 4000  \\
$1.400$ &  -0.021(5)    & 35.6(3.7)      &  0.0001(1)    &  4.356(6)    & 5000  \\
$1.500$ &  -0.024(7)    & 29.5(5.1)      &  0.0001(1)    &  4.26(1)     & 2200  \\
$1.600$ &  -0.005(5)    & 33.6(4.2)      &  0.0001(1)    &  4.194(9)    & 5850  \\
$1.700$ &  -0.005(5)    & 31.5(3.7)      &  0.0001(1)    &  4.17(2)     & 2400  \\
\end{tabular}
\caption{ Observables measured on a $16^4$ lattice with four-fermi coupling 
$\lambda = 1/G=1.4$}
\end{table}
\narrowtext

\begin{table}
\begin{tabular} {cddddd}
$\beta_g$ & $\sigma$ &  $\chi_{\sigma}$   &  $M$   &  $\chi_M$   &  $Trajectories$   \\ \hline
$0.200$ &   2.909(2)   &  5.35(75)    &  0.8480(3)     &  1.24(1)    & 650  \\
$0.300$ &   2.2450(18)  &  3.61(50)    &  0.8341(4)     &  1.39(1)    & 550  \\
$0.400$ &   1.8271(10)  &  3.75(51)    &  0.7967(4)     &  1.857(7)   & 2200  \\
$0.500$ &   1.5370(6)  &  3.35(35)    &  0.7378(4)     &  2.762(8)   & 12200  \\
$0.525$ &   1.4760(7)  &  2.76(35)    &  0.6824(3)     &  3.75(1)    & 9650  \\
$0.550$ &   1.4180(5)  &  2.64(34)    &  0.6162(3)     &  5.19(1)    & 10200  \\
$0.5625$ &  1.3839(7)  &  3.15(50)    &  0.0350(3)     &  31.6(2)    & 6650  \\
$0.575$ &   1.3592(6)  &  3.17(45)    &  0.0532(5)     &  32.8(1)    & 11200  \\
$0.5875$ &  1.3327(6)  &  2.75(45)    &  0.0351(3)     &  30.3(1)    & 8700  \\
$0.59375$ & 1.3180(5)  &  2.83(36)    &  0.0372(6)     &  27.5(2)    & 10600  \\
$0.600$ &   1.3086(6)  &  3.58(35)    &  0.085(1)      &  22.2(2)    & 6100  \\
$0.700$ &   1.1233(8)  &  2.34(55)    &  0.0198(2)     &  18.54(7)   & 1700  \\
$0.800$ &   0.9784(8)  &  3.85(45)    &  0.0150(2)     &  15.33(7)   & 3800  \\
$0.900$ &   0.8485(7)  &  3.05(48)    &  0.0149(2)     &  13.37(6)   & 2600  \\
\end{tabular}
\caption{ Observables measured on a $16^4$ lattice with gauge coupling
$\beta = 1.0$}
\end{table}
\narrowtext

\end{document}